\documentclass[useAMS,usenatbib]{mn2e}

\usepackage{graphicx}
\usepackage{psfig}
\def\kpc{\,{\rm kpc}}
\def\kms{\,{\rm km\,s^{-1}}}
\def\cmss{\,{\rm cm\,s}^{-2}}

\def\fracj#1#2{{\textstyle{#1\over#2}}}

\title[MOND in the Milky Way]{Modified Newtonian Dynamics in the Milky Way}
\author[Benoit Famaey and James Binney]{Benoit Famaey\thanks{E-mail:
famaey@thphys.ox.ac.uk; binney@thphys.ox.ac.uk} and James
Binney\footnotemark[1]\\
Rudolf Peierls Centre for Theoretical Physics, Department of  Physics, University of Oxford, 1 Keble Road, Oxford OX1 3NP, United Kingdom}
\begin{document}

\date{Accepted ... Received ... ; in original form ...}

\pagerange{\pageref{firstpage}--\pageref{lastpage}} \pubyear{2004}

\maketitle

\label{firstpage}

\begin{abstract}
Both microlensing surveys and radio-frequency observations of gas flow
imply that the inner Milky Way is completely dominated by baryons, contrary
to the predictions of standard cold dark matter (CDM) cosmology. We
investigate the predictions of the Modified Newtonian Dynamics (MOND)
formula for the Galaxy given the measured baryon distribution. Satisfactory
fits to the observationally determined terminal-velocity curve are obtained
for different choices of the MOND's interpolating function $\mu(x)$.
However, with simple analytical forms of $\mu(x)$, the local circular speed
$v_c(R_0)$ can be as large as $220\kms$ only for values of the parameter
$a_0$ that are excluded by observations of NGC 3198. Only a numerically
specified interpolating function can produce $v_c(R_0)=220\kms$, which is
therefore an upper limit in MOND, while the asymptotic velocity is predicted
to be $v_c(\infty)=170\pm5\kms$. The data are probably not consistent with
the functional form of $\mu(x)$ that has been explored as a toy model in the
framework of Bekenstein's covariant theory of gravity. 
\end{abstract}

\begin{keywords}
gravitation -- Galaxy: kinematics and dynamics
\end{keywords}

\section{\label{sec:intro}Introduction\protect\\}

Recent advances in the modelling of the Milky Way
indicate that, contrary to the predictions of cold dark matter (CDM)
cosmology \citep[][and references therein]{DiemandMS}, the inner Galaxy is
completely dominated by baryons. We ask whether current Galaxy models are
more compatible with the Modified Newtonian Dynamics (MOND).

The evidence for accelerating cosmic expansion can be interpreted either as
signalling the presence of a large amount of dark energy (quintessence), or
as an indication that Einstein's minimal theory of gravity must be modified
by adding a cosmological constant. Similarly, the evidence for flat galactic
rotation curves is conventionally interpreted as evidence for massive halos
of cold dark matter (CDM), but twenty years ago \cite{Mil} suggested that
flat rotation curves might signal the need to modify Newtonian dynamics in
regions where the acceleration is smaller than a critical value $a_0$.  This
proposal, dubbed MOND, was then refined by the introduction of a
non-relativistic field equation for the modified gravitational potential
\citep{BekensteinM}.  There is now a significant body of evidence that, for
whatever reason, there {\it is\/} a characteristic acceleration $a_0\sim
cH_0$ associated with galaxies \citep{SandersM}. Furthermore, MOND has a
remarkable ability to account for features in the phenomenology of galaxies
that were unknown when Milgrom introduced the theory.

Recent developments in the theory of gravity have added plausibility to the
case for modification of gravity rather than addition of exotic matter.
First, \cite{Bekenstein04} has presented a Lorentz-covariant theory of
gravity that has a MONDian behaviour  in the appropriate limit.
Second, it has become recognized \citep[e.g.][]{Gripaios04} that spontaneous
symmetry breaking in an effective field theory of gravity might well lead to
loss of Lorentz invariance of the type required by Bekenstein. A great deal
of work needs to be done to determine whether a theory such as that of
Bekenstein is compatible with observations of the cosmic microwave
background and large-scale structure \citep[e.g.][]{Sko05}. But knowledge that it {\it is\/}
possible to embed MOND within a physically acceptable dynamical framework,
and that this framework is not unattractive from the point of view of
mathematical physics, must make us take more seriously the phenomenological
successes of MOND.

On the other hand, there is growing evidence that the dark halos of
galaxies do not conform to CDM simulations: they are cuspy in neither
low-surface-brightness (e.g. Bosma 2004) nor
high-surface-brightness galaxies, the centers of which are completely
baryon-dominated (Gerhard et al. 2001, Gentile et al. 2004, Cappellari et
al. 2005). Moreover the shapes of the rotation curves are intimately
connected to the underlying stellar-light profile, just as modified gravity
predicts. In this paper, we study the particular case of our own Milky Way
galaxy, which is the best example of a baryon-dominated
high-surface-brightness galaxy. 

In Section 2 we review the evidence that the inner Galaxy is dominated by
baryons. In Section 3 we investigate the predictions of MOND for the
Galactic circular-speed curve and the vertical equilibrium of the Solar
neighbourhood. Section 4 sums up.

\section{The Basel model of the Milky Way\protect\\}

\cite{BG} (hereafter BG) present a model of the luminosity density
interior to the Sun.  Like its predecessor \citep{BinneyGS}, this model is
based on the COBE $L$-band photometry, but it incorporates constraints from the
slight difference in the mean apparent magnitude of red-clump stars on
either side of the Galactic centre, and from tracers of spiral arms
in the disk. Its bar is thinner than that in the model of Binney et al.\ (1997).

\cite{BEG} (hereafter BEG) determine the pattern speeds of the bar and the
spiral structure, and upgrade the BG model to a mass model by simulating the
flow of gas in potentials obtained by assigning spatially constant
mass-to-light ratios to the BG model.  Maps of the density of CO and HI in
the longitude-velocity plane [$(l,v)$ plane] are derived and compared with
the corresponding observational plots. BEG find that they can reproduce the
observed $(l,v)$ plots by assigning to the bar a pattern speed that agrees
with an independent determination from the kinematics of the solar
neighbourhood \citep{Dehnen00}, while assigning a significantly lower pattern speed to the spiral structure. The spiral structure is four-armed and its amplitude in the mass
density is larger by a factor $\sim1.5$ than its amplitude in the
near-infrared luminosity density. This standard model explains the entire rotation curve inside a
galactocentric radius of $\sim0.6R_0$ (where $R_0$ is the galactocentric
radius of the Sun) without any dark matter. However, at the solar
radius $R_0$, the model yields a circular speed that is $\sim40\kms$ too
low. To make up the balance, an axisymmetric quasi-isothermal dark halo with
a large core radius ($r_c=1.34R_0$) and a small central density is added,
and the mass-to-light ratio of the luminous component is lowered by less
than 10\% compared to the standard model. 

The mass model developed by BEG is strongly supported by the results of
surveys for microlensing events in fields towards the Galactic centre.  Once
the mass-to-light ratio of the BG model has been determined, a map of the
microlensing optical depth for clump stars can be produced without any
further assumptions, and BG present such a map. In units of $10^{-6}$, the
EROS collaboration report an optical depth $\tau_6=0.94\pm0.3$ at
$(l,b)=(2.5^\circ,-4^\circ)$ \citep{Afonso} while BG predicted $\tau_6=1.2$
at this location. The MACHO collaboration report
$\tau_6=2.17^{+0.47}_{-0.38}$ at $(l,b)=(1.5^\circ,-2.68^\circ)$
\citep{Popowski04}, while BG predicted $\tau_6=2.4$ here. Any
transfer of matter from the stellar component to the dark halo would impair
this excellent agreement.  Still further evidence of the correctness of the
Basel model has been provided by an $N$-body model that self-consistently
generates the mass density inferred by BEG \citep{BDG}. This model, which has
no free parameters, successfuly predicts the proper motions of stars in
various bulge fields. When a very plausible model of the mass function of
bulge and disk stars is adopted, it also reproduces the distribution
determined by the MACHO collaboration of the duration of individual
microlensing events. Thus the Basel mass model, and its negligible
contribution from CDM to the inner Galaxy, has been very thoroughly
validated.

\begin{figure}
   \centering
   \psfig{file=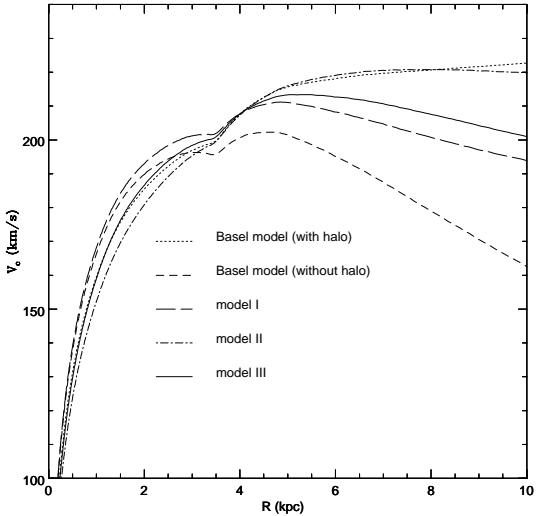,width=.9\hsize}
      \caption{Circular-speed curves for the Milky Way. Dotted and
	  short-dashed curves: Basel model with (dotted) and without
(short-dashed) a quasi-isothermal dark halo.  Long-dashed and full curves
are MONDian curves for $a_0= 1.2 \times 10^{-8} \cmss$ and either standard
(short-dashed) or simple (full) interpolating functions.  The dot-dashed
curve is the MONDian circular-speed curve with the standard interpolating function and $a_0= 3.9 \times 10^{-8} \cmss$.
\label{rotcurve}
       } 
   \end{figure}

The Basel group is now producing an upgraded version of the Basel model
(Englmaier, private communication) and the calculations that follow are
based on this new version. The model without dark matter (short-dashed
curve of Fig.~\ref{rotcurve}) is 12\% heavier
than the luminous part of the model with a dark halo (dotted curve of Fig.~\ref{rotcurve}). In the latter, the
quasi-isothermal dark halo has a core radius $r_c=10.7$~kpc and an asymptotic
velocity $v_\infty=235\kms$. It is assumed that $R_0 =8$~kpc and that
$v_c(R_0)=220\kms$. The model with dark
matter does not match the bumps in the rotation curve as quite as well as
the standard model without dark matter does.

\section{\label{sec:mond}The Milky Way in MOND\protect\\}
The halo introduced by BEG is not cuspy like the halos predicted by CDM
cosmology (Diemand et al. 2004, and references therein). A cuspy halo could
not have been added: indeed, what drives BEG to assign to baryons
essentially all matter within the corotation radius of the bar, and to
enhance the amplitude of the spiral structure outside the bar, is the size
of the non-circular velocities that are apparent in the observed $(l,v)$
plots: the dark halo is assumed to be axisymmetric, so if much mass is shifted
from the baryons to the halo, the non-axisymmetric component of the overall
potential is weakened, and the non-circular velocities are predicted to be
too small. Given that the structure of the BEG halo is unexpected in the CDM
theory, we now ask whether it can be eliminated by assuming that MOND is the
correct theory of gravity. A full answer to this question requires extensive
numerical work to solve the non-linear field equation \citep{BekensteinM}
for the modified potential generated by the Galaxy, including contributions
from the bar and spiral structure. However, the non-axisymmetry is important
only at $R\la\frac12R_0$, where the effects of MOND are small, which guarantees that the problem arising in the dark matter framework will not arise in MOND. Therefore,
as in standard Newtonian theory, the first step is to study the
circular-speed curve that follows from the axisymmetric component of
the density distribution.

\begin{figure}
   \centering
   \psfig{file=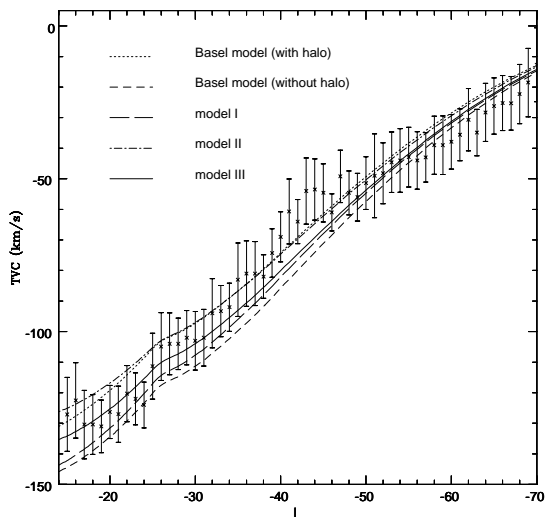,width=.9\hsize}
      \caption{Terminal velocity curves in the fourth quadrant for the same
models as in Fig.~1, compared with the data of Kerr et al.\ (1986). The terminal lign-of-sight velocities are defined by $v_t(l) = {\rm sign}(l) v_c(R_0\sin l)-v_c(R_0)\sin l$. 
\label{tvc}
	} 
   \end{figure}

\subsection{The rotation curve}

In cylindrical or spherical symmetry the gravitational force per unit mass
${\bf K}_{\rm MOND}$ predicted by MOND in the Galactic plane is related to
the corresponding Newtonian force per unit mass ${\bf K}_{\rm Newton}$ by
Milgrom's formula \citep{Mil}:
 \begin{equation}\label{mueq}
        \mu(K_{{\rm MOND}}/a_0) \, {\bf K}_{{\rm MOND}} = {\bf K}_{{\rm Newton}} ,
\end{equation}
 where the {\it interpolating function\/} $\mu$ runs smoothly from
$\mu(x)=x$ at $x\ll1$ to $\mu(x)=1$ at $x\gg1$.  In a flat axisymmetric
disk, Milgrom's formula is only exact if MOND is viewed as a modification of
inertia \citep{Mil94} rather than a modification of gravity. However, using
the field equation for the modified gravitational potential, \cite{BrM95} have shown that equation
(\ref{mueq}) holds everywhere outside Kuzmin discs and disc-plus-bulge
generalizations of them, while it provides a very good approximation for
exponential discs. As a consequence, this formula has been used to fit the
rotation curves of an impressive list of external galaxies \citep{SandersM}.
It is thus worthwhile to test the ability of this formula to fit the rotation
curve of the Milky Way.

Once the values of $a_0$ and of the mass-to-light ratio $\Upsilon$ are
known, equation (\ref{mueq}) predicts the force field for each choice of
interpolating function. From a sample of external galaxies with high quality
rotation curves \cite{Begeman} derived $a_0 = 1.2\pm 0.27 \times 10^{-8}
\cmss$ using the `standard' interpolating function $\mu(x)=x/\sqrt{1+x^2}$.
Retaining the mass-to-light ratio of the Basel model without dark matter, and
using the standard interpolating function yields the long-dashed circular-speed
curve  shown on Fig.~\ref{rotcurve} (Model I). The circular speed is
then $207\kms$ at $R=\frac{1}{2}R_0$, exactly as in the Basel model with dark halo, and $200\kms$ at $R_0$. Thus making gravity MONDian with the standard interpolating function reduces the deficit in
$v_c(R_0)$ from $40\kms$ to $20\kms$, but does not eliminate it.

\begin{table}
 \centering
 \begin{minipage}{75mm}
  \caption{For the Basel model with and without dark matter, and our 4 MONDian models, the columns display respectively the form of the interpolating function $\mu$, the value of $a_0$ in units of $10^{-8}
\cmss$, the value of the $L$-band mass-to-light ratio $\Upsilon_L$ in
$M_\odot/L_\odot$ ($\Upsilon_L = \xi^2$, where $\xi$ is the scaling
constant for the rotation curve in BEG), the circular speed at the solar
radius $R_0$ and at infinity in $\kms$, and finally the value $\mu(x_0)$
where $x_0 = v_c^2(R_0)/(R_0 \, a_0)$. The values are displayed for models
with $R_0 = 8$~kpc.}
  \begin{tabular}{@{}lcccccc@{}}
  \hline
    & $\mu(x)$ & $a_0$ & $\Upsilon_L$ & $v_c(R_0)$ & $v_\infty$ & $\mu(x_0)$\\
 \hline
DM & 1 & 0 & 1.08 & 220 & 235 & 1\\
no-DM& 1 & 0 & 1.21 & 180 & 0 & 1\\
I & $x/\sqrt{1+x^2}$ & 1.2 & 1.21 & 200 & 175 & 0.8\\
II &  $x/\sqrt{1+x^2}$& 3.4 & 0.91 & 220 & 210 & 0.4\\
III & $x/(1+x)$ & 1.2 & 0.95 & 208 & 165 & 0.6\\
IV & fit & 1.2 & 1.08 & 220 & 170 & 0.6\\
\hline
\end{tabular}
\end{minipage}
\end{table}

However, the value of $v_c(R_0)$ is hard to determine and a lower value
than $220\kms$ is
not excluded \cite[e.g.][]{Ol98}.  The primary observable is the terminal
velocity $v_t(l)$ at each longitude $l$:
\begin{equation}
v_t(l) = {\rm sign}(l) v_c(R_0\sin l)-v_c(R_0)\sin l.
\end{equation}
From $v_t(l)$ it is easy to
determine $v_c(R)$ if one knows $R_0$ and $v_c(R_0)$, but neither parameter
is known with precision. Given this uncertainty, it is important to
understand the predictions of each model for the run of $v_t(l)$ (see Fig.~\ref{tvc}). When $v_c(R_0) = 220\kms$ is assumed, as it was
by the Basel group, the baryonic Basel model yields an excellent fit to the
data at $|l|\la40^\circ$, but the curve is not self-consistent because it
does not pass through zero at $|l|=90^\circ$. The short-dashed curve in
Fig.~\ref{tvc} shows that when we self-consistently adopt $v_c(R_0)=180\kms$
(see Table 1), in the fourth quadrant the terminal velocities near
$l=-20^\circ$ become marginally too negative and from there rise too steeply
as $l\to-90^\circ$. However, this self-consistent baryonic Newtonian model
is not clearly incompatible with the measured terminal velocities. 

Our MONDian Model I (long-dashed curve in Fig.~2) reduces the discrepancy
with the therminal velocity data, but still fits the data less well than the
Basel model with dark matter (dotted curve in Fig.~2) because
$v_c(R_0)=200\kms$ is low. If we enforce $v_c(R_0)=220\kms$ simply by
increasing $\Upsilon$, we make $v_c$ too large at $R\approx\fracj12R_0$.  A
better fit can be obtained by adjusting both $a_0$ and $\Upsilon$.  To
increase $v_c(R_0)$ we increase $a_0$ and thus trigger a MONDian correction
at smaller radii.  This in turn obliges us to decrease $\Upsilon$ in order
to keep a velocity of $207\kms$ at $R=\fracj12R_0$, exactly as in the Basel
model with dark matter.  This decrease in $\Upsilon$ requires a further
increase in $a_0$ to retain $v_c(R_0)=220\kms$, with the result that by the
time we have achieved a satisfactory fit to the rotation curve at
$R=\fracj12R_0$ and $R=R_0$, $a_0$ has increased significantly.  The
dot-dashed curves in Figs.~\ref{rotcurve} and \ref{tvc} shows the fits
obtained when $a_0= 3.4 \times 10^{-8} \cmss$ and $\Upsilon_L = 0.91 \,
M_\odot/L_\odot$ (Model II). This model predicts an asymptotic circular
speed of $210\kms$ that is higher than those predicted by the conventional
value of $a_0$ (see Table 1).

\begin{figure}
   \centering
   \psfig{file=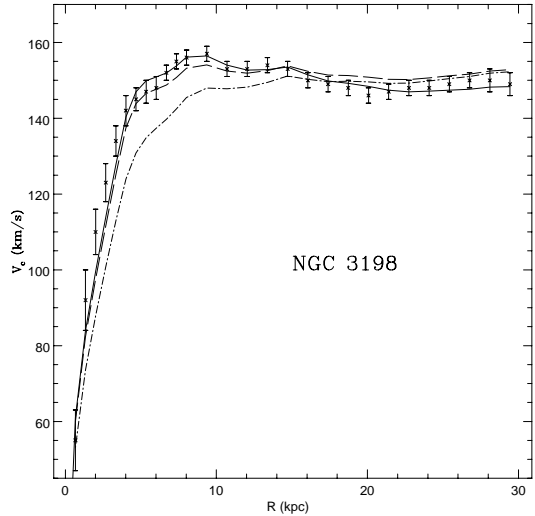,width=.9\hsize}
      \caption{Fits to the rotation curve of NGC 3198 using the parameters
of Model I (long-dashed), II (dot-dashed) and III (continuous) of Table 1.
Models I and III are compatible with NGC 3198, while Model II is
incompatible. This means that the value of $a_0$ used in Model II with the
classical interpolating function is too high to account for a fit of NGC
3198. On the other hand the interpolating function of Model III provides an
even better fit than the interpolating function used in Model II.
\label{extcurves}
       } 
   \end{figure}

The asymptotic circular speed of the Galaxy is not well determined
\citep{DehnenB,WilkinsonE}. By contrast, in external galaxies with extended
HI, the behaviour of $v_c$ at large $R$ is strongly constrained by
observations.  Extended rotation curves have been obtained for many galaxies
with the aim of probing dark halos \citep[e.g.][]{Begeman,Broeils,Gentile}.
To be specific, we focus here on the case of NGC 3198, which was carefully
studied by \cite{Begeman}; this system is the textbook example of a galaxy
with an extended flat rotation curve because it has a large apparent
diameter and a velocity field that shows the gas to be confined to a plane
and following accurately circular orbits.  Fig.~\ref{extcurves} shows that
when one fits the rotation curve of NGC 3198 with $a_0=3.4 \times 10^{-8}
\cmss$, the decrease in $\Upsilon$ that is required to match the measured
asymptotic velocity results in $v_c$ being too low at small radii. 

Consequently, the data for NGC 3198 rule out the large value of $a_0$
required by Model II. In principle we can make Model II compatible with the
standard value $a_0=1.2\times10^{-8}\cmss$ by rescaling it: the rotation
curves of the models listed in Table 1 are invariant if we change the value
of $R_0$ from $8\kpc$ and then multiply $\Upsilon$ by $R_0/(8\,{\rm kpc})$
and $a_0$ by $(8\,{\rm kpc})/R_0$. However, to obtain
$a_0=1.2\times10^{-8}\cmss$ by rescaling $R_0$, we would have to set $R_0
\simeq 15\kpc$, which is excluded by a wealth of data on the scale size of
the Milky Way. 

From this we conclude that, with the standard interpolating function, MOND cannot fit with
perfect accuracy the rotation curves of both the Basel model and NGC
3198. This conclusion is consistent with the work of \cite{Gentile}, who
concluded that by varying only $\Upsilon$ it is not possible to obtain
entirely satisfactory fits to the rotation curves of five spiral galaxies.

We therefore consider an alternative `simple' interpolating function, namely
 \begin{equation}
\mu(x)={x\over1+x}.
\end{equation}
This function provides a less sudden transition from the Newtonian to the
MONDian regime than does the standard function. The continuous curve in
Fig.~\ref{extcurves} shows that the simple
interpolating function together with the conventional value $a_0=1.2 \times
10^{-8} \cmss$ yield a extremely good fit to the rotation curve
of NGC 3198. 

The full curves in Figs.~\ref{rotcurve} and \ref{tvc} show the corresponding
fits to the Basel model (Model III).  We now have $v_c(R_0)=208\kms$.
Although this value lies very close to the conventional value, we find it is
impossible to push $v_c(R_0)$ up to $220\kms$ by increasing $a_0$, because
we then have to decrease the mass-to-light ratio radically in order to fit
the inner rotation curve [$v_c(0.5 \, R_0)=207\kms$], making $v_c(R_0)$ too
low again. The transition from Newtonian to MONDian physics provided by the
simple interpolating function is insufficiently abrupt. Slightly changing
the value of $R_0$ does not eliminate the problem. However,
Fig.\ref{tvc} shows that the fit of Model III to the terminal-velocity curve
is extremely satisfactory. This simple interpolating function thus fits the
data for both NGC 3198 and the Milky Way. However, it does not exactly reproduce the
run of $v_c(R)$ of the Basel model; to achieve this, one needs a
more complex form of the interpolating function.

\subsection{The interpolating function\protect\\}

In low-surface-brightness (LSB) galaxies, $x<1$ at all radii, with the
consequence that the measured acceleration $v_c^2/R$ is approximately equal
to $[a_0\Upsilon GL(R)/R^2]^{1/2}$. It follows that these galaxies constrain
only the product $a_0\Upsilon$ and leave the interpolating function
unconstrained. In high-surface-brightness (HSB) galaxies such as
NGC 3198, $x$ ranges from values greater than unity down to small
values. Hence with these objects 
we can break the degeneracy between $a_0$ and $\Upsilon$, but, as
Fig.~\ref{extcurves} illustrates, the large
gradients in rotation curves at small radii, combined with the poor spatial
resolution of obsevations in the 21-cm line, leave considerable degeneracy
between $a_0$ and $\Upsilon$, and constrain the interpolating function only
weakly. The Basel model is strongly constrained at small radii, and it is
pertinent to probe the extent to which it constrains the interpolating
function in equation (\ref{mueq}).

Imagine an ideal galaxy in which both the asymptotically flat rotation curve
and the luminosity distribution are known with precision. Then with $x =
v_{c}^2/(Ra_0)$ we can write
 \begin{equation}\label{fitmu}
\mu(x) = \Upsilon \frac{f(a_0 x)}{a_0 x},
\end{equation}
 where $f(v_c^2/R)$ is obtained from the Newtonian equation for the force
per unit mass with the mass-density $\rho(\bf x)$ replaced by the
luminosity-density $j(\bf x)$.
From data at large $x$ (corresponding to  the central parts of the
galaxy), where $\mu(x)\simeq1$ and $f(y)\propto y$, we can read off
$\Upsilon$. At large radii (small $x$) we have $\mu(x)\simeq x$ and
$f(y)\propto y^2$, so we can determine $a_0$. 
Once $\Upsilon$ and $a_0$ are known, we can completely determine $\mu(x)$ from
equation (\ref{fitmu}).

As an example, we treat the circular-speed curve of the Basel model plus dark halo
as a perfect set of data. From the centre we find $\Upsilon_L = 1.08 \,
M_\odot/L_\odot$, while the asymptotic velocity $v_\infty=235\kms$ implies
$a_0= 4.47 \times 10^{-8} \cmss$. This large value of $a_0$ suggests that
the model's asymptotic velocity is too high.  If one could show
that the Milky Way's asymptotic circular speed really is this high, MOND would
have been falsified.

\begin{figure}
   \centering
   \psfig{file=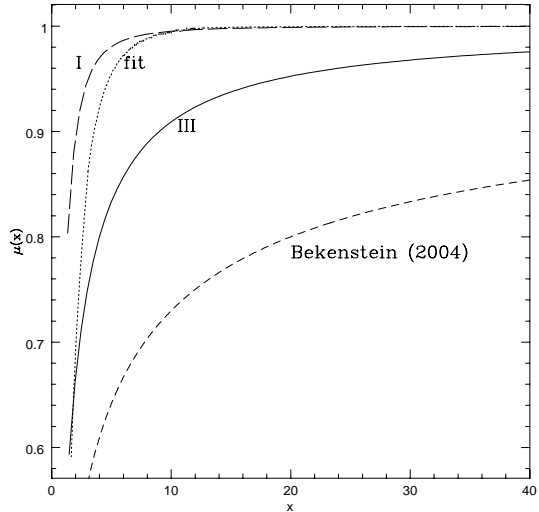,width=.9\hsize}
      \caption{Fit (dotted line) to the interpolating function of MOND if
one takes the Basel model as an ideal set of data from $R=0$ until $R_0$,
and if one assumes an asymptotic velocity of $170\kms$. The long-dashed and
continuous lines correspond to the interpolating function of Models I and III,
respectively. The short-dashed line is the interpolating function explored
by Bekenstein (2004)
 \label{fitcurve}
}
   \end{figure}

However, if we combine $\Upsilon_L = 1.08 \, M_\odot/L_\odot$ with the
conventional value $a_0=1.2 \times 10^{-8} \cmss$, we infer an
asymptotic velocity $170\kms$, and can read off the form of the
interpolating function that perfectly fits the Basel model at $R\le R_0$ (Model IV).
The dotted curve in Fig.~\ref{fitcurve} shows this function, which
transitions smoothly from $x/(1+x)$ at $x\le1$ to $x/\sqrt{1+x^2}$ at
$x\ge10$.

Any successful underlying theory for MOND must predict an interpolating
function that lies between the curves labelled I and III in
Fig.~\ref{fitcurve}. \cite{Bekenstein04} introduced the TeVeS theory of
gravitation, in which the physical metric depends on the Einstein metric, a
vector field, and a scalar field. The dynamics of the scalar field is
controlled by a Lagrangian density involving an undetermined function $F$
that yields a MONDian behaviour for the potential in the nonrelativistic
limit. Adopting a computationally tractable form for $F$, Bekenstein shows
for the case of spherical symmetry that Eq.~(1) holds with $\mu$ replaced by
a function that depends on both the Newtonian acceleration and the gradient
of the scalar field. For accelerations that are not large compared to $a_0$,
the MONDian and Newtonian accelerations are then related by Eq.~(1) with the
interpolating function
 \begin{equation}
\mu(x) = {\sqrt{1 + 4x} -1\over\sqrt{1 + 4x} +1},
\end{equation}
 which is shown as the dashed curve in Fig.~\ref{fitcurve}. Since this
curve lies far below the dotted curve at all values of $x$, the
correction to Newtonian gravity to which this function gives rise is
much too big to be compatible with the dynamics of the Milky Way (see
Table 1).  Of course, since the Galaxy is not spherical, the
interpolating function (5) does not apply to it, and even the concept of an interpolating function cannot be used for the
Galaxy in TeVeS. However, it is not obvious that when non-spherical
geometry is taken into account, Bekenstein's toy $F$ will yield a
sufficiently rapid transition from the MONDian to the Newtonian
regime. Hence, groups that are endeavouring to determine the
implications of TeVeS for large-scale structure should be cautious
about conclusions obtained using Bekenstein's toy $F$, and especially
not disregard this gravitational theory if predictions happen to
disagree with observations when using this particular $F$.

\subsection{The vertical equilibrium}

We now examine the consistency of MOND with the vertical dynamics of the
solar neighbourhood. This is in principle a complex problem, because out of
the plane it is not a priori clear that the Newtonian and modified-gravity
accelerations will be parallel, as Milgrom's formula (\ref{mueq}) implies.
However, equation (\ref{mueq}) is known to hold outside Kuzmin and
exponential disks, as well as in cylindrical symmetry, so it may be expected to provide a good first
approximation in the Milky Way.

At $R_0$ (assumed here to be 8~kpc), the surface density of baryons is
$\Sigma_0 = 53 \, {\rm M}_\odot \, {\rm pc}^{-2}$ \citep{HolmbergF}. An
infinite sheet with this surface density generates a Newtonian vertical
acceleration $K_{z \, {\rm Newton}} \simeq \pi G \Sigma_0 \simeq 0.25 a_0$,
for $a_0 = 1.2 \times 10^{-8} \cmss$.
The radial acceleration is $v_c^2/R_0\simeq 6.5 \times K_{z \, {\rm
Newton}}$, which implies that the modification of the vertical force is
dominated by the external field effect. Thus $K_{z \, {\rm MOND}}\simeq K_{z
\, {\rm Newton}}/\mu(v_c^2/R_0 \, a_0)$.
Observations show that  $1.1\,$kpc above the plane, the
vertical force is larger than that arising from the baryons in Newtonian
theory by a factor $\simeq7/5$ \citep{KuijkenG, HolmbergF}. It follows that
MOND requires that
 \begin{equation}
\mu(v_c^2/R_0 \, a_0) \simeq 5/7\simeq0.71.
\end{equation}

The extreme right column of Table 1 shows that, given the significant
uncertainties, this constraint is satisfied by the standard, simple and
fitted interpolating functions when $a_0=1.2\times10^{-8}\cmss$. The choice
$a_0=3.4\times10^{-8}\cmss$ used in Model II places the solar neighbourhood
so deeply into the MOND regime that the predicted large correction to the
vertical force is excluded by the observations.

\section{Discussion\protect\\}

By combining near-infrared photometry and simulations of gas flow in the
plane, and without invoking dark matter in the inner $5\kpc$, the Basel
group has developed an extremely successful model of the Milky Way that
accounts for the structure of $(l,v)$ plots for CO and HI, for the proper
motions of bulge stars, for the microlensing optical depth towards bulge
fields, and for the observed distribution of the durations of microlensing
events. Given the number of these checks, there can be little doubt that we
now really do know the distribution of baryons inside the solar radius. For
no other galaxy do we have information of comparable quality.

It is far from clear that the Basel model is compatible with the predictions
of CDM. In the light of early indications that baryons dominate the inner
Galaxy, an attempt was made to build models that minimize the dark halo
consistent with constraints from simulations of clustering CDM
\citep{KlypinZS}. In all these models, however, CDM contributes
significantly to the density at $\sim3\,$kpc from the Galactic centre where
the Basel model requires the density to be almost entirely invested in
stars. To investigate this matter further, we need a CDM-inspired model that
includes a stellar bar and reproduces the photometry of the Galaxy. It would
be of great interest to calculate the predictions of such a model for
microlensing surveys and the $(l,v)$ diagrams of CO and HI. If those
predictions disagree with observations, the CDM paradigm would be strongly
weakened. \cite{Gentile} have shown that the paradigm encounters similar
difficulties reproducing the rotation curves of a sample of five external
galaxies. 

In MOND, there are two free parameters ($\Upsilon$ and $a_0$, the latter a
constant of nature) and one free function ($\mu$, to be fixed by an
underlying theory of MOND). Once those quantities are known, the
gravitational force field is completely determined by the baryon
distribution. We have investigated what circular-speed curves MOND predicts from the distribution of baryons in the Basel model. A full answer to this
question requires extensive numerical work, but a good approximation to the
truth can be obtained from Milgrom's formula (\ref{mueq}). For the value of
$a_0$ that has been determined from observations of external galaxies, we
encounter difficulty making $v_c(R_0)$ as large as the value, $220\kms$,
assumed by the Basel model. When the standard interpolating function
$\mu(x)$ is used, satisfaction of this condition requires
$a_0=3.4\times10^{-8}\cmss$ (Model II), a value which is incompatible with
observations of NGC 3198. Keeping the conventional value
$a_0=1.2\times10^{-8}\cmss$ (Model I) yields $v_c(R_0)=200\kms$ implying
slightly too low values for the terminal velocities. Interestingly, Gentile
et al. (2004) have also found that MOND cannot account for the rotation curves of their sample of spiral galaxies, at least with that value of $a_0$ and
that choice of interpolating function. However, in the Milky Way, a very good
fit to the terminal velocity curve (Model III) can be obtained if the
interpolating function (3) is used. It would be of great interest to test
the ability of this interpolating function to fit the rotation curves of
external galaxies in the sample of Gentile et al. (2004). We then showed
how a perfect set of data could determine $\Upsilon$, $a_0$ and the
interpolating function $\mu(x)$, and displayed the results (Model IV) that
would follow if data were precisely those predicted by the Basel model at
$R\le R_0$.  

In the absence of dark matter, we conclude that the Galaxy's asymptotic
velocity must be smaller than the value, $235\kms$, implied by the dark halo
of the Basel model (see Table 1); MOND predicts it to be $\sim 170 \pm 5
\kms$. The fitted, numerically specified interpolating function of Model IV
produces a local circular speed of $220\kms$, but lower values are obtained
for the standard and simple interpolating functions of Models I and III,
which suggests that $220\kms$ is an upper limit for $v_c(R_0)$. We have also
shown that it is unlikely that the interpolating function explored by
Bekenstein (2004) as a toy model can account for the dynamics of the Milky
Way. We finally showed that the three models with
$a_0=1.2\times10^{-8}\cmss$ (Models I, III and IV) are compatible with what
is known about the vertical force at $R_0=8$ kpc.

It can thus be argued that, with a suitable choice of MOND's interpolating
function, and a realistic choice of mass-to-light ratio, the Milky Way can
be added to the long list of galaxies for which Milgrom's formula
(\ref{mueq}) is successful in predicting the rotation curve from the baryon
distribution. The next step towards exploring the agreement between the data
and the predictions MONDian gravity makes from the Basel model involves
using a potential solver \citep[e.g.][]{BrM99} for the nonlinear
Bekenstein--Milgrom equation to determine the MOND force generated by the
non-axisymmetric components of the Basel model with the parameters and
interpolating functions of our Models I, III and IV. Another test for the
relevance of MOND as an alternative to dark matter in the Milky Way will be
provided by the measurement of the velocity dispersions in distant globular
clusters orbiting the Galaxy in the outer halo, well into the deep-MOND
regime (Baumgardt, Grebel \& Kroupa 2005).

\section*{Acknowledgements}

We thank P. Englmaier and O. Gerhard for kindly providing us the circular-speed curves on which this paper is based. We also thank the {\it Fondation Wiener-Anspach} (Belgium) for its financial support.

\bsp

\label{lastpage}

\end{document}